\RequirePackage{lineno}
\documentclass[aps,twocolumn,showpacs,byrevtex,prl,reprint]{revtex4-1}

\usepackage{graphicx}
\usepackage{dcolumn}
\usepackage{bm}
\usepackage{rotating}
\usepackage{epstopdf}
\usepackage{color}
\usepackage{verbatim} 
\usepackage{multirow}
\usepackage[abs]{overpic}
\usepackage{amsmath}
\usepackage{amssymb}
\usepackage{xspace}
\usepackage[titletoc]{appendix}
\usepackage{float}

\newcommand{\PreserveBackslash}[1]{\let\temp=\\#1\let\\=\temp}
\newcolumntype{C}[1]{>{\PreserveBackslash\centering}p{#1}}
\newcolumntype{R}[1]{>{\PreserveBackslash\raggedleft}p{#1}}
\newcolumntype{L}[1]{>{\PreserveBackslash\raggedright}p{#1}}

\newcommand{\EE}{e^+e^-}

\newcommand{\too}{\rightarrow}

\begin{document}
\graphicspath{{figure/}}
\DeclareGraphicsExtensions{.eps,.png,.ps}
\title{\boldmath First study of reaction $\Xi^{0}n\too\Xi^{-}p$ using $\Xi^0$-nucleus scattering at an electron-positron collider}
\author{
  \begin{small}
    \begin{center}
      M.~Ablikim$^{1}$, M.~N.~Achasov$^{13,b}$, P.~Adlarson$^{75}$, R.~Aliberti$^{36}$, A.~Amoroso$^{74A,74C}$, M.~R.~An$^{40}$, Q.~An$^{71,58}$, Y.~Bai$^{57}$, O.~Bakina$^{37}$, I.~Balossino$^{30A}$
% ...
\\
\vspace{0.2cm}
(BESIII Collaboration)\\
\vspace{0.2cm} {\it
$^{1}$ Institute of High Energy Physics, Beijing 100049, People's Republic of China\\
% ...
$^{13}$ G.I. Budker Institute of Nuclear Physics SB RAS (BINP), Novosibirsk 630090, Russia\\
% ...
$^{30}$ INFN Sezione di Ferrara, (A)INFN Sezione di Ferrara, I-44122, Ferrara, Italy; (B)University of Ferrara,  I-44122, Ferrara, Italy\\
% ...
$^{36}$ Johannes Gutenberg University of Mainz, Johann-Joachim-Becher-Weg 45, D-55099 Mainz, Germany\\
$^{37}$ Joint Institute for Nuclear Research, 141980 Dubna, Moscow region, Russia\\
% ...
$^{40}$ Liaoning Normal University, Dalian 116029, People's Republic of China\\
% ...
$^{57}$ Southeast University, Nanjing 211100, People's Republic of China\\
$^{58}$ State Key Laboratory of Particle Detection and Electronics, Beijing 100049, Hefei 230026, People's Republic of China\\
% ...
$^{71}$ University of Science and Technology of China, Hefei 230026, People's Republic of China\\
% ...
$^{74}$ University of Turin and INFN, (A)University of Turin, I-10125, Turin, Italy; (B)University of Eastern Piedmont, I-15121, Alessandria, Italy; (C)INFN, I-10125, Turin, Italy\\
$^{75}$ Uppsala University, Box 516, SE-75120 Uppsala, Sweden\\
% ...
$^{b}$ Also at the Novosibirsk State University, Novosibirsk, 630090, Russia\\
% ...
      }\end{center}
    \vspace{0.4cm}
\end{small}
}
\affiliation{}


\begin{abstract}
Using $(1.0087\pm0.0044)\times10^{10}$ $J/\psi$ events collected with the BESIII detector at the BEPCII storage ring, the process $\Xi^{0}n\too\Xi^{-}p$ is studied, where the $\Xi^0$ baryon is produced in the process $J/\psi\too\Xi^0\bar{\Xi}^0$ and the neutron is a component of the $^9\rm{Be}$, $^{12}\rm{C}$ and $^{197}\rm{Au}$ nuclei in the beam pipe. A clear signal is observed with a statistical significance of $7.1\sigma$. The cross section of the reaction $\Xi^0+{^9\rm{Be}}\too\Xi^-+p+{^8\rm{Be}}$ is determined to be $\sigma(\Xi^0+{^9\rm{Be}}\too\Xi^-+p+{^8\rm{Be}})=(22.1\pm5.3_{\rm{stat}}\pm4.5_{\rm{sys}})$~mb at the $\Xi^0$ momentum of $0.818$~GeV/$c$, where the first uncertainty is statistical and the second is systematic. No significant $H$-dibaryon signal is observed in the $\Xi^-p$ final state. This is the first study of hyperon-nucleon interactions in electron-positron collisions and opens up a new direction for such research.
\end{abstract}

\maketitle

Scattering experiments of high energy particle beams bombarding target materials have been of great significance for studying the inner structure of matter and the fundamental interactions~\cite{introduction1, protonobs, neutronobs}. Charged long-lived particle beams such as $\pi^{\pm}$ and $K^{\pm}$ can be easily produced, and relevant research results have been very rich. On the other hand, due to significantly shorter lifetimes and higher masses, particle beams of hyperons, such as $\Lambda$, $\Sigma$, or $\Xi$, are more difficult to produce and corresponding experiments are rare, although measurements of these beams bombarding target materials are crucial for understanding non-perturbative quantum chromodynamics (QCD).

The experimental study on the interaction between hyperons and different target materials began in the 1960s and has lasted for more than half a century~\cite{introduction2, introduction3, introduction4, introduction5, introduction6, introduction7, introduction8, introduction9, introduction10, introduction11}. However, the intensities of hyperon beams produced by these experiments are relatively low and relevant experimental measurements are very scarce. After stagnating for decades, in 2021 and 2022, the CLAS and J-PARC E40 Collaborations reported the latest results of $\Lambda/\Sigma$ and nucleon interaction respectively, including the reactions $\Lambda p\too\Lambda p$~\cite{introduction12}, $\Sigma^{-}p\too\Sigma^{-}p$~\cite{introduction13}, $\Sigma^{+}p\too\Sigma^{+}p$~\cite{introduction14} and $\Sigma^{-}p\too\Lambda n$~\cite{introduction15}. More research on hyperon--nucleon interaction is still strongly needed. Compared with other hyperons, relevant experimental measurements on $\Xi$--nucleon interaction are even more limited. Only a few events were observed for each reaction~\cite{introduction5, introduction6, introduction7, introduction8, introduction9, introduction10, introduction11}. The interaction of $\Xi$ and nucleons has been studied in some theoretical models, such as the constituent quark model~\cite{introduction16, introduction17, introduction18}, the meson-exchange picture~\cite{introduction19}, and the chiral effective field theory approach~\cite{introduction20, introduction21, introduction22}. More experimental measurements are needed to constrain the theoretical models, which can greatly promote research in this field.

The study of $\Xi$--nucleon interaction also can be used to search for the $H$-dibaryon, which has strangeness $-2$ and valence quark structure $uuddss$. This $H$-dibaryon was first predicted according to the bag model in the 1970s~\cite{introduction23, introduction24}. Later studies by two lattice QCD groups also predicted the existence of the $H$-dibaryon~\cite{introduction25, introduction26, introduction27}. Although the $H$-dibaryon has been searched for by many experiments, no convincing signal has been found so far~\cite{introduction28, introduction29, introduction30, introduction31, introduction32, introduction33, introduction34, introduction35}. The $H$-dibaryon can also be searched for in $\Xi$--nucleon scattering processes, for example, in the $\Xi^-p$ final state in the process $\Xi^{0}n\too\Xi^{-}p$. Especially, Refs.~\cite{introduction23, introduction24} predict an $H$-dibaryon may appear as a bound state of $\Sigma\Sigma$ decaying strongly into $\Xi N$ or $\Lambda\Lambda$, where $N$ represents $n$ or $p$. Furthermore, the study of hyperon--nucleon interactions is important to understand the role of hyperons in dense neutron-star matter, to determine the equation of state (EoS) of nuclear matter at supersaturation densities and to understand the so-called ``hyperon puzzle" of neutron stars~\cite{introduction36, introduction37, introduction38}. The study of $\Xi$--nucleon interaction is also helpful to understand the formation of $\Xi$ hypernuclei, on which experimental information is very scarce~\cite{introduction39, introduction40, introduction41}.

\begin{figure}[htbp]
\begin{center}
\begin{overpic}[width=0.32\textwidth]{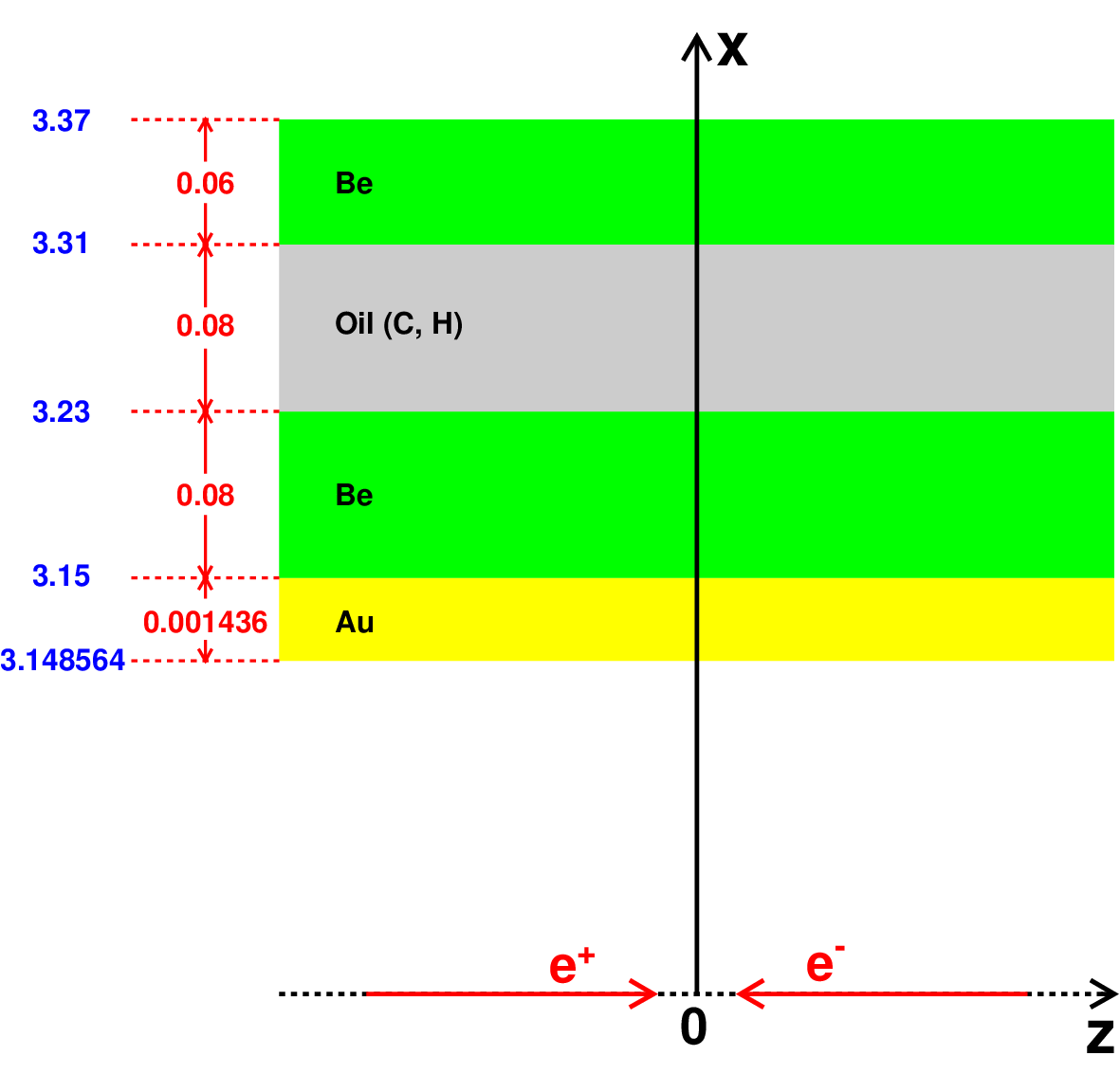}
\end{overpic}
\caption{Schematic diagram of the beam pipe, the length units are centimeter (cm). The $z$-axis is the symmetry axis of the MDC, and the $x$-axis is perpendicular to the $e^+e^-$ beam direction.}
\label{fig:beampipe}
\end{center}
\end{figure}

The BESIII detector records symmetric $e^+e^-$ collisions at the BEPCII collider~\cite{bepcii}. Details of the BESIII detector can be found in Ref.~\cite{besiii}. With a sample of $(1.0087\pm0.0044)\times10^{10}$ $J/\psi$ events collected by the BESIII detector~\cite{totalnumber}, an intense monoenergetic $\Xi$ baryon can be produced by the decay $J/\psi\too\Xi^{0}\bar{\Xi}^{0}$. The $\Xi$ baryon can interact with the material in the beam pipe adjacent to the $e^+e^-$ beam, providing a novel source to study the $\Xi$--nucleon interaction~\cite{introduction42, introduction43}. The material of the beam pipe is composed of gold ($^{197}\rm{Au}$), beryllium ($^{9}\rm{Be}$) and oil $(^{12}\rm{C}:$$^{1}\rm{H}$$=1:2.13)$, as shown in Fig.~\ref{fig:beampipe}, with more details in Ref.~\cite{besiii}.

In this Letter, we describe a study of the reaction $\Xi^{0}n\too\Xi^{-}p$, where the $\Xi^0$ baryon is produced in the process $J/\psi\too\Xi^0\bar{\Xi}^0$, and the neutron is a component of the $^9\rm{Be}$, $^{12}\rm{C}$ and $^{197}\rm{Au}$ nuclei in the beam pipe. This is the first study of hyperon--nucleon interaction at an electron--positron collider. The cross section of the reaction $\Xi^0+{^9\rm{Be}}\too\Xi^-+p+{^8\rm{Be}}$ is also determined. Since the momentum of the monoenergetic, incident $\Xi^0$ is relatively high ($P_{\Xi^0}=0.818$~GeV/$c$), its interaction with atomic nuclei tends to be a direct nuclear reaction. It is assumed that a $\Xi^0$ reacts with a neutron in the $^9\rm{Be}$ directly in the reaction, and that afterwards a residual nucleus $^8\rm{Be}$, a $\Xi^-$ and a proton are left over. To determine the cross section of $\Xi^0+{^9\rm{Be}}\too\Xi^-+p+{^8\rm{Be}}$ from the composite material, the reaction is assumed to be a pure surface process.

In this analysis, simulated data samples that are produced with a {\sc{Geant4}}-based~\cite{geant4} Monte Carlo (MC) package, which includes the geometric description of the BESIII detector~\cite{display} and the detector response, are used to determine detection efficiencies and to estimate backgrounds. The inclusive MC sample includes both the production of the $J/\psi$ resonance and the continuum process incorporated in {\sc kkmc}~\cite{KKMC}. All particle decays are modeled with {\sc evtgen}~\cite{ref:evtgen} using branching fractions either taken from the Particle Data Group (PDG)~\cite{pdg}, where available, or otherwise estimated with {\sc lundcharm}~\cite{ref:lundcharm}. Final state radiation (FSR) from charged final state particles is incorporated using the {\sc photos} package~\cite{photos}.

The signal process considered in this analysis is $J/\psi\too\Xi^0\bar{\Xi}^0$, $\Xi^{0}n\too\Xi^{-}p$, $\Xi^-\too\Lambda\pi^-$, $\Lambda\too p\pi^-$, $\bar{\Xi}^0\too\bar{\Lambda}\pi^0$, $\bar{\Lambda}\too \bar{p}\pi^+$, $\pi^0\too\gamma\gamma$. In order to determine the detection efficiency, $1.0\times10^6$ signal MC events are simulated, with the angular distribution of $J/\psi\too\Xi^0\bar{\Xi}^0$ generated according to the measurement in Ref.~\cite{alpha}.  We simulate the reaction process $\Xi^{0}n\too\Xi^{-}p$ assuming the neutron to be free, regardless of its Fermi-momentum. Since the momentum of the monoenergetic, incident $\Xi^0$ is much greater than the Fermi-momentum, this approximation is reasonable. The effect of this approximation is considered in the systematic uncertainty evaluation. Since the distribution of the $\Xi^-p$ invariant mass, $M(\Xi^-p)$, is almost flat in data from $2.26$ to $2.40$~GeV/$c^2$, the mass of free neutron in MC simulation is tuned to change the center-of-mass energy of the reaction system to make the $M(\Xi^-p)$ distribution consistent between data and MC. The angular distribution of the reaction process is generated using an isotropic phase-space distribution.

Charged tracks detected in the multilayer drift chamber (MDC) are required to be within a polar angle ($\theta$) range of $|\cos\theta|<0.93$, where $\theta$ is defined with respect to the $z$-axis taken to be the symmetry axis of the MDC. Photon candidates are identified using showers in the electromagnetic calorimeter (EMC). The deposited energy of each shower must be more than 25~MeV in the barrel region $(|\cos\theta|<0.8)$ and more than 50~MeV in the end cap region $(0.86<|\cos\theta|<0.92)$. To exclude showers that originate from charged tracks, the angle enclosed by the EMC shower and the position of the closest charged track at the EMC must be greater than 10 degrees as measured from the interaction point. To suppress electronic noise and showers unrelated to the event, the difference between the EMC time and the event start time is required to be within $[0, 700]$~ns. Particle identification for charged tracks combines measurements of the energy deposited in the MDC (d$E$/d$x$) and the flight time in the time-of-flight system (TOF) to form likelihoods $\mathcal{L}(h)$ $(h=p, K, \pi)$ for each hadron $h$ hypothesis. Tracks are identified as protons when the proton hypothesis has the greatest likelihood $(\mathcal{L}(p)>\mathcal{L}(\pi)$ and $\mathcal{L}(p)>\mathcal{L}(K))$, while tracks are identified as pions when the pion hypothesis has the greatest likelihood $(\mathcal{L}(\pi)>\mathcal{L}(K)$ and $\mathcal{L}(\pi)>\mathcal{L}(p))$.

Since the final state of the signal process is $pp\pi^-\pi^-\bar{p}\pi^{+}\gamma\gamma$, candidate events must have six charged tracks with zero net charge and at least two photon candidates. We require that there are two $p$, two $\pi^-$, one $\bar{p}$ and one $\pi^+$. For the decay $\bar{\Xi}^0\too\bar{\Lambda}\pi^0$ with $\bar{\Lambda}\too \bar{p}\pi^+$, we perform a vertex fit to the $\bar{p}\pi^+$ combination, and the $\bar{\Lambda}$ signal region is defined as $|M(\bar{p}\pi^+)-m_{\bar{\Lambda}}|<0.003$~GeV/$c^{2}$, where $m_{\bar{\Lambda}}$ is the nominal mass of the $\bar{\Lambda}$. In this Letter, all nominal masses are taken from PDG~\cite{pdg}. The invariant mass of the two photons is required to be in the $\pi^0$ mass window $[0.11, 0.15]$~GeV/$c^2$, and the invariant mass of the two photons is constrained to the nominal mass of the $\pi^0$ using a 1C kinematic fit. If there is more than one $\pi^0$ candidate in an event, only the one with the minimum value of $|M(\bar{\Lambda}\pi^0)-m_{\bar{\Xi}^0}|$ is retained. The $\bar{\Xi}^0$ signal region is defined as $-0.015$~GeV/$c^{2}<(M(\bar{\Lambda}\pi^0)-m_{\bar{\Xi}^0})<0.010$~GeV/$c^{2}$, where $m_{\bar{\Xi}^0}$ is the $\bar{\Xi}^0$ nominal mass. For the reaction $\Xi^{0}n\too\Xi^{-}p$ with subsequent decay $\Xi^-\too\Lambda\pi^-$, we first perform the vertex fit of $\Lambda$ by considering all $p\pi^-$ combinations. The $p\pi^-$ combination with the smallest value of $|M(p\pi^-)-m_{\Lambda}|$, where $m_{\Lambda}$ is the $\Lambda$ nominal mass, is taken as $\Lambda$ candidate. The $\Lambda$ signal region is also defined as $|M(p\pi^-)-m_{\Lambda}|<0.003$~GeV/$c^{2}$. Then, a vertex fit of $\Xi^-$ is performed for the combination of the $\Lambda$ and the remaining $\pi^-$. Finally, a vertex fit is performed for the combination of the $\Xi^-$ and the remaining $p$.

To select the signal events of $J/\psi\too\Xi^0\bar{\Xi}^0$, the invariant mass of the system recoiling against the $\bar{\Xi}^0$, $M_{\text{recoil}}(\bar{\Xi}^0)$, is required to be in the $\Xi^0$ signal region, defined as $[1.295, 1.325]$ GeV/$c^{2}$, where $M_{\text{recoil}}(\bar{\Xi}^0)\equiv\sqrt{E^2_{\text{beam}}-|\vec{p}_{\bar{\Xi}^0}c|^2}/c^2$, $E_{\text{beam}}$ is the $e^+$ or $e^-$ beam energy for data, and $\vec{p}_{\bar{\Xi}^0}$ is the measured momentum of the $\bar{\Xi}^0$ candidate in the $\EE$ rest frame. The main background is $J/\psi\too\Xi^0\bar{\Xi}^0$, $\Xi^0\too\Lambda\pi^0$, $\bar{\Xi}^0\too\bar{\Lambda}\pi^0$. To suppress this background, the recoil mass of $\bar{\Xi}^0\Lambda$, $M_{\text{recoil}}(\bar{\Xi}^0\Lambda)$, is obtained from the four-momenta of the initial $e^+e^-$ system and the $\bar{\Xi}^0$ and $\Lambda$ candidates. $M_{\text{recoil}}(\bar{\Xi}^0\Lambda)$ should be around the nominal $\pi^0$ mass for this background, so we require $M_{\text{recoil}}(\bar{\Xi}^0\Lambda)<0$~GeV/$c^2$ to remove these events.

\begin{figure}[htbp]
\begin{center}
\begin{overpic}[width=0.32\textwidth]{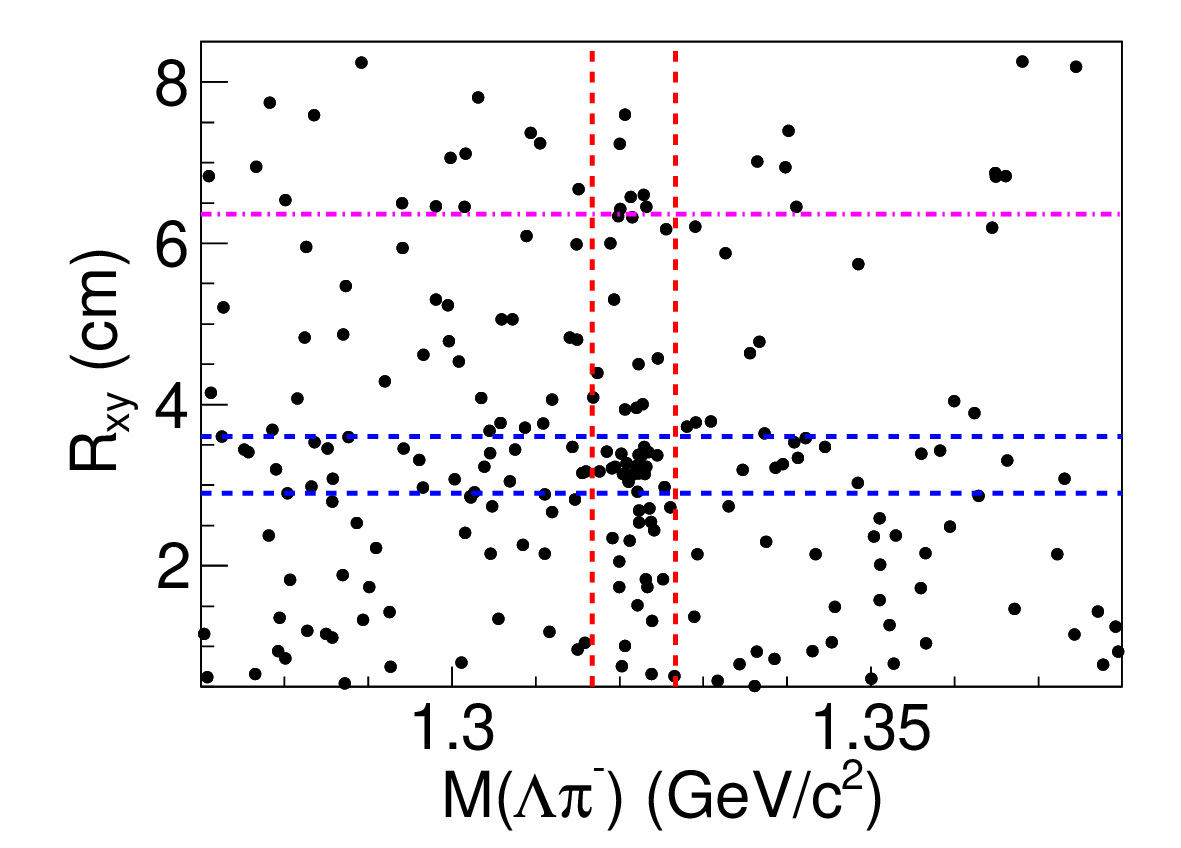}
\end{overpic}
\caption{Distribution of $R_{xy}$ versus $M(\Lambda\pi^-)$ for data. The blue horizontal dashed lines denote the beam pipe region, the pink horizontal dashed-dotted line denotes the position of inner wall of MDC, and the red vertical dashed line marks the $\Xi^-$ signal region.}
\label{fig:scatter_xirxy}
\end{center}
\end{figure}

The distribution of $R_{xy}$ versus the invariant mass $M(\Lambda\pi^-)$ of data is shown in Fig.~\ref{fig:scatter_xirxy}, where $R_{xy}$ is the distance from the reconstructed $\Xi^-p$ vertex to the $z$-axis. The beam pipe signal region is defined as $[2.9, 3.6]$~cm, taking into account the detector resolution.  This requirement also removes the influence from additional supporting materials. Clear enhancements are seen in the beam pipe and $\Xi^-$ signal regions, due to $\Xi^0$ interactions with material in the beam pipe producing $\Xi^{-}$ via the process $\Xi^{0}n\too\Xi^{-}p$.  A cluster of events can be seen in the inner wall of MDC region, defined as $[6.0, 6.8]$~cm, but the signal is not statistically significant.

\begin{figure}[htbp]
\begin{center}
\begin{overpic}[width=0.32\textwidth]{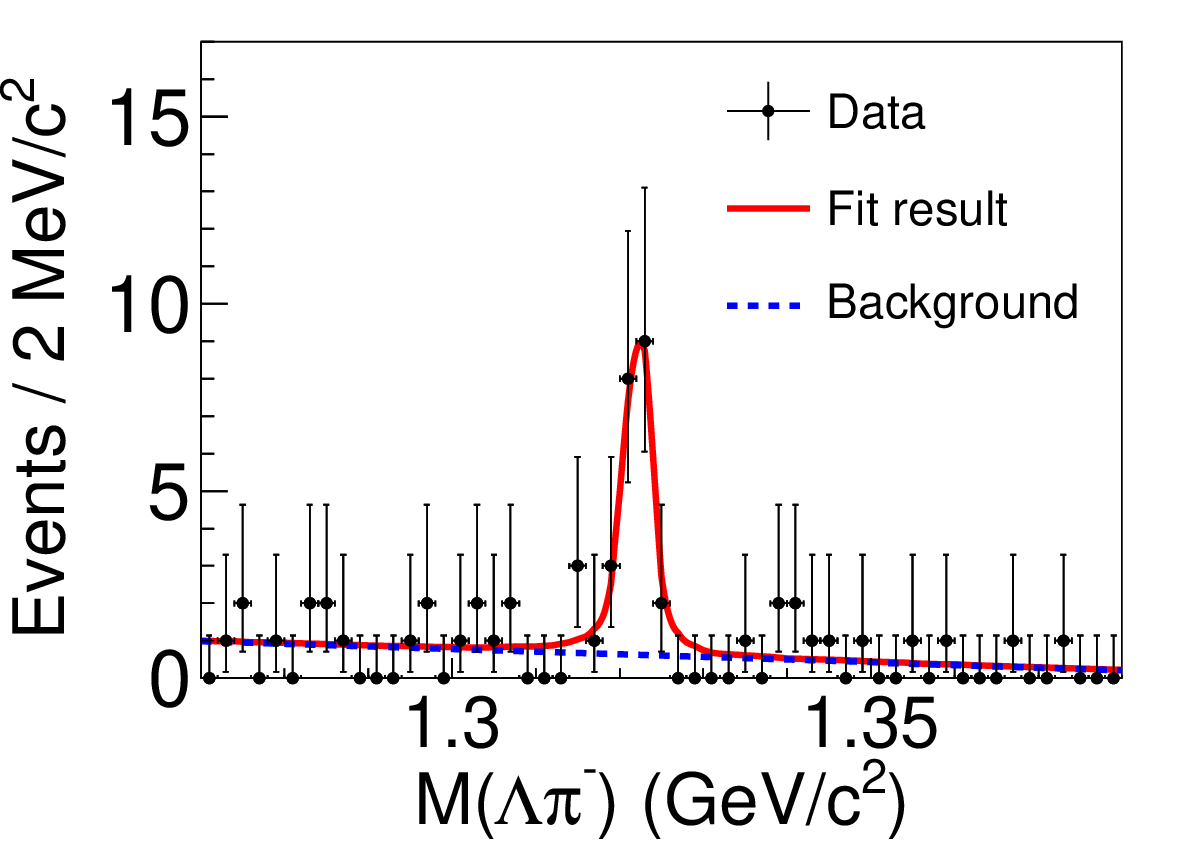}
\end{overpic}
\caption{Distribution of $M(\Lambda\pi^-)$ in data (dots with error bars). The red solid curve is the total fit result and the blue dashed curve is the background component.}
\label{fig:fit_result}
\end{center}
\end{figure}

Figure~\ref{fig:fit_result} shows the $M(\Lambda\pi^-)$ distribution from data after final event selection. A clear $\Xi^-$ signal is observed, corresponding to the reaction $\Xi^{0}n\too\Xi^{-}p$. A detailed study of the $J/\psi$ inclusive MC sample indicates that there is no peaking background contribution in the $\Xi^-$ signal region. Additionally, no significant peak is found in beam pipe sideband events from data. To determine the signal yield, an unbinned maximum likelihood fit is performed to the $M(\Lambda\pi^-)$ distribution. We use the MC-determined shape to describe the $\Xi^-$ signal, where the yield acts as a free fit parameter. The background is described by a linear function with the number of events and the slope as free parameters. The fit result is shown in Fig.~\ref{fig:fit_result}; the $\Xi^-$ signal yield returned by the fit is $N^{\rm{sig}}=22.9\pm5.5$. The statistical significance is determined to be $7.1\sigma$ by comparing the likelihood values for the fits with and without the $\Xi^-$ signal and taking the change of the number of degrees-of-freedom into account.

Since the beam pipe is composed of layers of composite material, as shown in Fig.~\ref{fig:beampipe}, the cross section of the reaction between $\Xi^0$ baryons and $^{9}\rm{Be}$ nuclei $\sigma(\Xi^0+{^9\rm{Be}}\too\Xi^-+p+{^8\rm{Be}})$ is extracted using
\begin{equation}
    \sigma(\Xi^0+{^9\rm{Be}}\too\Xi^-+p+{^8\rm{Be}}) = \frac{N^{\rm{sig}}} {\epsilon \mathcal{B} \mathcal{L}_{\rm{eff}} },
\end{equation}
where $\epsilon$ is the selection efficiency, $\mathcal{B}$ is the product of the branching ratios of all intermediate resonances, defined as $\mathcal{B}\equiv\mathcal{B}(\bar{\Xi}^0\too\bar{\Lambda}\pi^0)\mathcal{B}(\bar{\Lambda}\too\bar{p}\pi^+)\mathcal{B}(\pi^0\too\gamma\gamma)
\mathcal{B}(\Xi^-\too\Lambda\pi^-)\mathcal{B}(\Lambda\too p\pi^-)$, and $\mathcal{L}_{\rm{eff}}$ is the effective luminosity of the $\Xi^0$ flux produced from $J/\psi\too\Xi^0\bar{\Xi}^0$ and the distribution of target materials, as shown in the following formula:
\begin{widetext}
\begin{equation}
    \mathcal{L}_{\rm{eff}} = \frac{\it{N}_{\it{J}/\psi}\mathcal{B}_{\it{J}/\psi}}{\rm{2}+\frac{\rm{2}}{\rm{3}}\alpha} \int_{a}^{b}\int_{\rm{0}}^{\pi} (\rm{1}+\alpha \rm{cos}^2\theta) \it{e}^{-\frac{x}{\rm{sin}\theta \it{\beta\gamma L}}}N(x)C(x) \rm{d}\theta \rm{d}\it{x}.
\end{equation}
\end{widetext}

In the formula for the effective luminosity, the angular distribution of the $\Xi^0$ flux, the attenuation of the $\Xi^0$ flux, the number of target nuclei, and the weight of different target materials are considered in turn. $N_{J/\psi}$ is the number of $J/\psi$ events~\cite{totalnumber}, $\mathcal{B}_{J/\psi}$ is the branching fraction of $J/\psi\too\Xi^0\bar{\Xi}^0$, $\alpha$ is the parameter of the angular distribution of $J/\psi\too\Xi^0\bar{\Xi}^0$~\cite{alpha}, $\beta\gamma\equiv \frac{\sqrt{E_{\rm{beam}}^{\rm{2}}-m_{\rm{\Xi^0}}^{\rm{2}}c^{\rm{4}}}}{m_{\rm{\Xi^0}}c^{\rm{2}}}$ is the ratio of the momentum and the mass of the $\Xi^0$, $L\equiv c\tau$ is the product of the light speed and the mean lifetime of the $\Xi^0$, $N(x)$ is the number of target nuclei per unit volume, $a$ and $b$ are the distances from the inner surface and outer surface of the beam pipe to the $z$-axis, $\theta$ and $x$ are the angle and distance to the $z$-axis. The beam pipe can be regarded as infinitely long with respect to the product $\beta\gamma L$ of $\Xi^0$. $C(x)$ is the cross section ratio relative to $\sigma(\Xi^0+{^9\rm{Be}}\too\Xi^-+p+{^8\rm{Be}})$, where we assume the reaction is dominated by the interaction of a $\Xi^0$ baryon with a single neutron on the $^9\rm{Be}$ nucleus surface~\cite{ratio1, ratio2, ratio3, ratio4, ratio5}. The derivation of the formula can be found in Section I of the Supplemental Material, and the relevant parameters are listed in Table~\ref{tab:result}. The corresponding cross sections are determined to be $\sigma(\Xi^0+{^9\rm{Be}}\too\Xi^-+p+{^8\rm{Be}})=(22.1\pm5.3_{\rm{stat}}\pm4.5_{\rm{sys}})$~mb, or $\sigma(\Xi^0+{^{12}\rm{C}}\too\Xi^-+p+{^{11}\rm{C}})=(24.1\pm5.8_{\rm{stat}}\pm4.6_{\rm{sys}})$~mb, which are not independent and are derived from the same basic quantities.

\begin{table}[htbp]
\begin{center}
\caption{Input parameters for the cross section calculation using Eq.(1). The nominal values of $C(x)$  are obtained based on the pure surface process assumption, and the values in brackets are obtained based on the assumption, that the cross section is proportional to the number of neutrons in the nucleus.}
\label{tab:result}
\begin{tabular}{cc}
  \hline
  \hline
  Parameter & Result \\
  \hline
  $N^{\rm{sig}}$ & $22.9\pm5.5$ \\
  $\epsilon$     & $1.873\%$ \\
  $\mathcal{B}$  & $(40.114\pm0.444)\%$~\cite{pdg} \\
  $N_{J/\psi}$   & $(1.0087\pm0.0044)\times10^{10}$~\cite{totalnumber} \\
  $\mathcal{B}_{J/\psi}$ & $(0.117\pm0.004)\%$~\cite{pdg} \\
  $\alpha$       & $0.514\pm0.016$~\cite{alpha} \\
  $L$            & $(8.69\pm0.27)$~cm~\cite{pdg} \\
  $E_{\rm{beam}}$& $1.5485$~GeV \\
  $m_{\Xi^0}$    & $(1.31486\pm0.00020)$~GeV/$c^2$~\cite{pdg}\\
  $a$            & $3.148564$~cm~\cite{besiii} \\
  $b$            & $3.37$~cm~\cite{besiii} \\
  $N(x)$ & $\begin{cases}
           5.91\times10^{22}~\rm{cm^{-3}}, \ \ 3.148564\leq \it{x}\leq \rm{3.15}~\rm{cm} \\
           1.24\times10^{23}~\rm{cm^{-3}}, \ \ 3.15< \it{x}\leq \rm{3.23}~\rm{cm} \\
           3.45\times10^{22}~\rm{cm^{-3}}, \ \ 3.23< \it{x}\leq \rm{3.31}~\rm{cm} \\
           1.24\times10^{23}~\rm{cm^{-3}}, \ \ 3.31< \it{x}\leq \rm{3.37}~\rm{cm} \\
           \end{cases}$ \\
  $C(x)$ & $\begin{cases}
           8.437 \ (23.6), \ \ \ \ \ \ \ \ \ 3.148564\leq \it{x}\leq \rm{3.15}~\rm{cm} \\
           1.000 \ (1.00), \ \ \ \ \ \ \ \ \ 3.15< \it{x}\leq \rm{3.23}~\rm{cm} \\
           1.090 \ (1.20), \ \ \ \ \ \ \ \ \ 3.23< \it{x}\leq \rm{3.31}~\rm{cm} \\
           1.000 \ (1.00), \ \ \ \ \ \ \ \ \ 3.31< \it{x}\leq \rm{3.37}~\rm{cm} \\
           \end{cases}$ \\
  \hline
  \hline
\end{tabular}
\end{center}
\end{table}

\begin{figure}[htbp]
\begin{center}
\begin{overpic}[width=0.23\textwidth]{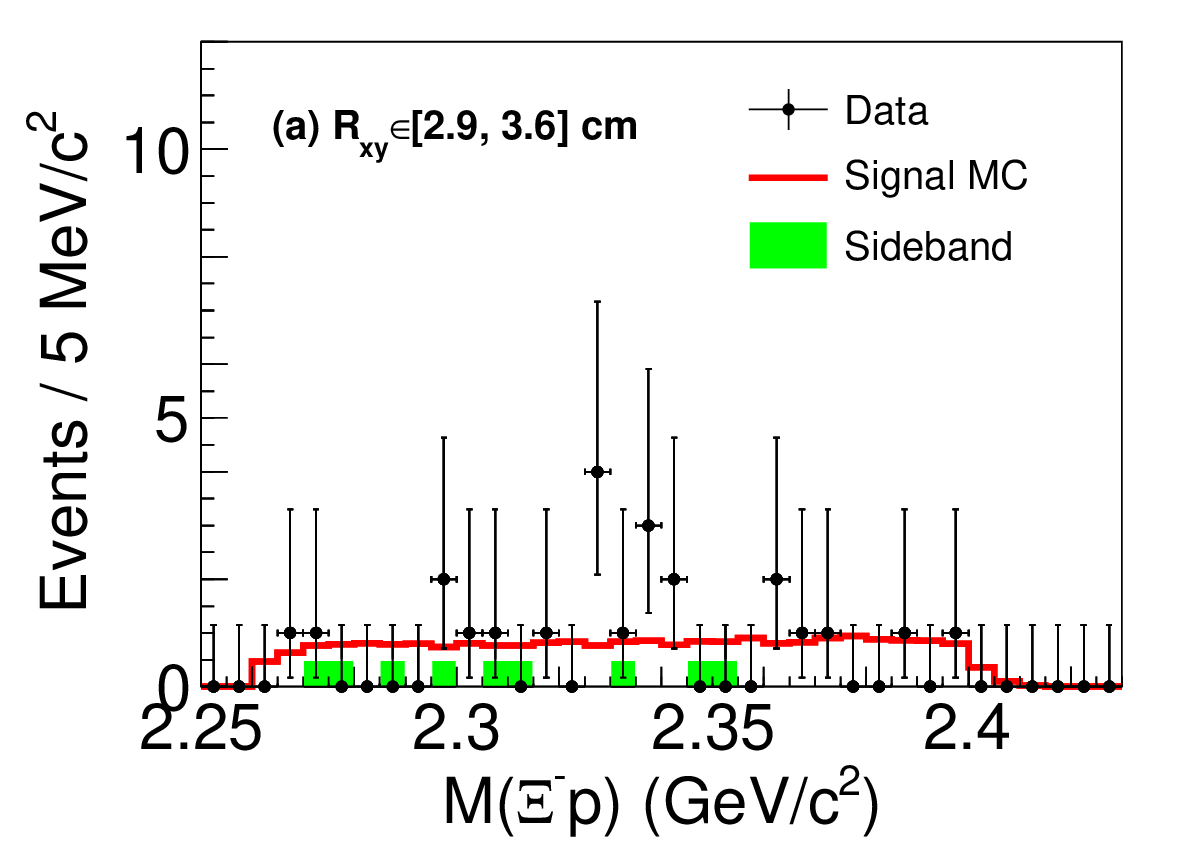}
\end{overpic}
\begin{overpic}[width=0.23\textwidth]{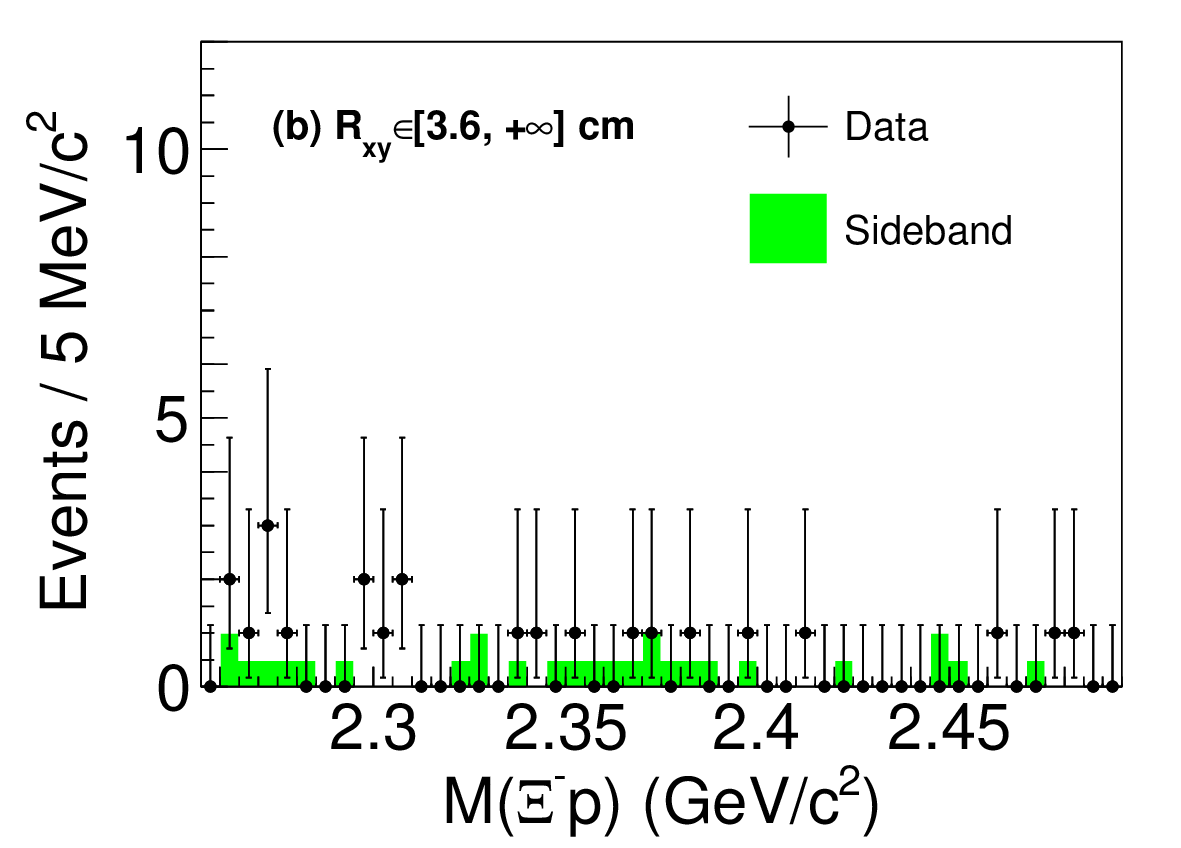}
\end{overpic}
\caption{Distributions of $M(\Xi^- p)$ for data in $2.9<R_{xy}<3.6$~cm (a) and $R_{xy}>3.6$~cm (b). The green shaded histograms correspond to the normalized events from the $\Xi^-$ sideband region, and the red line corresponds to the signal MC distribution that is normalized by the total number of events for data.}
\label{fig:fig_mH}
\end{center}
\end{figure}

We also search for $H$-dibaryon signals in the $\Xi^-p$ final state. The mean lifetime of the $H$-dibaryon is unknown, it may decay in the beam pipe region, or after flying some distance. Figure~\ref{fig:fig_mH} shows the $M(\Xi^- p)$ distributions for selected $\Xi^-$ signal events inside or outside the beam pipe region. Based on the available statistics, we do not see any obvious peaks in the two $M(\Xi^- p)$ distributions, so no significant short-lifetime or long-lifetime $H$-dibaryon signal is observed in the process $\Xi^{0}n\too\Xi^{-}p$ with $^{9}\rm{Be}$, $^{12}\rm{C}$ and $^{197}\rm{Au}$.

The sources of systematic uncertainties related to the determined cross sections come from the tracking efficiency (6.0\%)~\cite{track}, photon efficiency (2.0\%), PID efficiency (6.0\%), track number requirement (3.0\%), mass windows (7.8\%), $R_{xy}$ requirement (6.6\%), $M_{\text{recoil}}(\bar{\Xi}^0\Lambda)$ requirement (4.3\%), $(\Xi^-+p)$ momentum (10.0\%), $M(\Xi^-p)$ distribution (0.6\%), angular distribution of $\Xi^{0}n\too\Xi^{-}p$ (1.0\%), MC statistics (0.7\%), efficiency curve parameterization (0.5\%), fit procedure (5.0\%), number of $J/\psi$ (0.4\%), branching fractions (3.6\%), angular distribution of $J/\psi\too\Xi^0\bar{\Xi}^0$ (0.1\%), $\Xi^0$ mean lifetime (2.7\%), $e^+e^-$ interaction point (2.7\%), and cross section ratios (7.2\% or 2.1\% for different nuclei). All systematic uncertainties are discussed in detail in Section II of the Supplemental Material. The total systematic uncertainties are 20.4\% and 19.2\% for $\sigma(\Xi^0+{^9\rm{Be}}\too\Xi^-+p+{^8\rm{Be}})$ and $\sigma(\Xi^0+{^{12}\rm{C}}\too\Xi^-+p+{^{11}\rm{C}})$, respectively.

In summary, using $(1.0087\pm0.0044)\times10^{10}$ $J/\psi$ events collected with the BESIII detector operating at the BEPCII storage ring, the reaction $\Xi^{0}n\too\Xi^{-}p$ is observed with a statistical significance of $7.1\sigma$, where $\Xi^0$ is from the process $J/\psi\too\Xi^0\bar{\Xi}^0$ and $n$ is from materials in the beam pipe. The cross section of the reaction $\Xi^0+{^9\rm{Be}}\too\Xi^-+p+{^8\rm{Be}}$ at the momentum of the $\Xi^0$ of $P_{\Xi^0}=0.818$~GeV/$c$ is determined to be $\sigma(\Xi^0+{^9\rm{Be}}\too\Xi^-+p+{^8\rm{Be}})=(22.1\pm5.3_{\rm{stat}}\pm4.5_{\rm{sys}})$~mb, where the first uncertainty is statistical and the second is systematic. If the effective number of reaction neutrons in a $^9\rm{Be}$ nucleus is taken as $3$~\cite{introduction11}, the cross section of $\Xi^{0}n\too\Xi^{-}p$ for a single neutron is determined to be $\sigma(\Xi^{0}n\too\Xi^{-}p)=(7.4\pm1.8_{\rm{stat}}\pm1.5_{\rm{sys}})$~mb, consistent with theoretical predictions in Refs.~\cite{introduction18, introduction21, introduction22}. Furthermore, we do not observe any significant $H$-dibaryon signal in the $\Xi^-p$ final state for this reaction process.

This work is the first study of hyperon--nucleon interaction in electron--positron collisions, and opens up a new direction for such research. Other hyperons and nucleon interactions can also be studied, such as $\Lambda$ and $\Sigma$. Furthermore, we may be able to design targets of specific materials to study hyperon-nucleon interaction in future super tau-charm facilities~\cite{super1, super2}. With more statistics at that time, we can also study the momentum-dependent cross section distribution based on the hyperons from multi-body decays of $J/\psi$ or other charmonia.

The BESIII Collaboration thanks the staff of BEPCII and the IHEP computing center for their strong support. This work is supported in part by National Key R\&D Program of China under Contracts Nos. 2020YFA0406300, 2020YFA0406400; National Natural Science Foundation of China (NSFC) under Contracts Nos. 11635010, 11735014, 11835012, 11935015, 11935016, 11935018, 11961141012, 12022510, 12025502, 12035009, 12035013, 12061131003, 12192260, 12192261, 12192262, 12192263, 12192264, 12192265, 12150004; the Chinese Academy of Sciences (CAS) Large-Scale Scientific Facility Program; the CAS Center for Excellence in Particle Physics (CCEPP); Joint Large-Scale Scientific Facility Funds of the NSFC and CAS under Contract No. U1832207; CAS Key Research Program of Frontier Sciences under Contracts Nos. QYZDJ-SSW-SLH003, QYZDJ-SSW-SLH040; 100 Talents Program of CAS; The Institute of Nuclear and Particle Physics (INPAC) and Shanghai Key Laboratory for Particle Physics and Cosmology; ERC under Contract No. 758462; European Union's Horizon 2020 research and innovation programme under Marie Sklodowska-Curie grant agreement under Contract No. 894790; German Research Foundation DFG under Contracts Nos. 443159800, 455635585, Collaborative Research Center CRC 1044, FOR5327, GRK 2149; Istituto Nazionale di Fisica Nucleare, Italy; Ministry of Development of Turkey under Contract No. DPT2006K-120470; National Research Foundation of Korea under Contract No. NRF-2022R1A2C1092335; National Science and Technology fund of Mongolia; National Science Research and Innovation Fund (NSRF) via the Program Management Unit for Human Resources \& Institutional Development, Research and Innovation of Thailand under Contract No. B16F640076; Polish National Science Centre under Contract No. 2019/35/O/ST2/02907; The Royal Society, UK under Contract No. DH160214; The Swedish Research Council; U. S. Department of Energy under Contract No. DE-FG02-05ER41374.

\clearpage

\begin{widetext}

\textbf{\large
\boldmath Supplemental Material for ``First study of reaction $\Xi^{0}n\too\Xi^{-}p$ using $\Xi^0$-nucleus scattering at an electron-positron collider"
\unboldmath
}

\section{I. Derivation of the formula for the cross section $\sigma(\Xi^0+{^9\rm{Be}}\too\Xi^-+p+{^8\rm{Be}})$}

The formula for the cross section $\sigma(\Xi^0+{^9\rm{Be}}\too\Xi^-+p+{^8\rm{Be}})$ is:
\begin{equation}
    \sigma(\Xi^0+{^9\rm{Be}}\too\Xi^-+p+{^8\rm{Be}}) = \frac{N^{\rm{sig}}} {\epsilon \mathcal{B} \mathcal{L}_{\rm{eff}} },
\end{equation}
where $\epsilon$ is the selection efficiency, $\mathcal{B}$ is the product of the branching ratios of all intermediate resonances, defined as $\mathcal{B}\equiv\mathcal{B}(\bar{\Xi}^0\too\bar{\Lambda}\pi^0)\mathcal{B}(\bar{\Lambda}\too\bar{p}\pi^+)\mathcal{B}(\pi^0\too\gamma\gamma)
\mathcal{B}(\Xi^-\too\Lambda\pi^-)\mathcal{B}(\Lambda\too p\pi^-)$, and $\mathcal{L}_{\rm{eff}}$ is the effective luminosity of the $\Xi^0$ flux and target materials. In the formula for the effective luminosity, the angular distribution of the $\Xi^0$ flux, the attenuation of the $\Xi^0$ flux, the number of target nuclei, and the weight of different target materials are considered in turn. In the next step, each component in the formula will be introduced respectively.

The measured angular distribution of the process $J/\psi\too\Xi^0\bar{\Xi}^0$~\cite{alpha-supp} is:
\begin{equation}
    \frac{\rm{d}\it{N(\theta)}}{\rm{d}(\rm{cos}\theta)} \propto (1+\alpha \rm{cos}^2\theta),
\end{equation}
where $\alpha$ is the parameter of the angular distribution of $J/\psi\too\Xi^0\bar{\Xi}^0$, $\theta$ is the angle between the $\Xi^0$ and the beam direction, as shown in Fig.~\ref{fig:beampipe2}, and $\rm{cos}\theta$ is from $-1$ to $+1$. According to the integral formula:
\begin{equation}
    \int_{-1}^{1}\frac{\rm{d}\it{N(\theta)}}{\rm{d}(\rm{cos}\theta)}\rm{d}(\rm{cos}\theta) = \it{N}_{\it{J}/\psi}\mathcal{B}_{\it{J}/\psi},
\end{equation}
where $N_{J/\psi}$ is the number of $J/\psi$ events, $\mathcal{B}_{J/\psi}$ is the branching fraction of $J/\psi\too\Xi^0\bar{\Xi}^0$, we get
\begin{equation}
    \frac{\rm{d}\it{N(\theta)}}{\rm{d}(\rm{cos}\theta)} = \frac{\it{N}_{\it{J}/\psi}\mathcal{B}_{\it{J}/\psi}}{\int_{-1}^{1}(1+\alpha \rm{cos}^2\theta) \rm{d}(\rm{cos}\theta)} (1+\alpha \rm{cos}^2\theta) \\
    = \frac{\it{N}_{\it{J}/\psi}\mathcal{B}_{\it{J}/\psi}}{2+\frac{2}{3}\alpha} (1+\alpha \rm{cos}^2\theta).
\end{equation}
Therefore, the following formula can be obtained:
\begin{equation}
    \frac{\rm{d}\it{N(\theta)}}{\rm{d}\theta} = \frac{\it{N}_{\it{J}/\psi}\mathcal{B}_{\it{J}/\psi}}{2+\frac{2}{3}\alpha} (1+\alpha \rm{cos}^2\theta)\rm{sin}\theta,
\end{equation}
where $\theta$ goes from $0$ to $\pi$.

\begin{figure}[htbp]
\begin{center}
\begin{overpic}[width=0.45\textwidth]{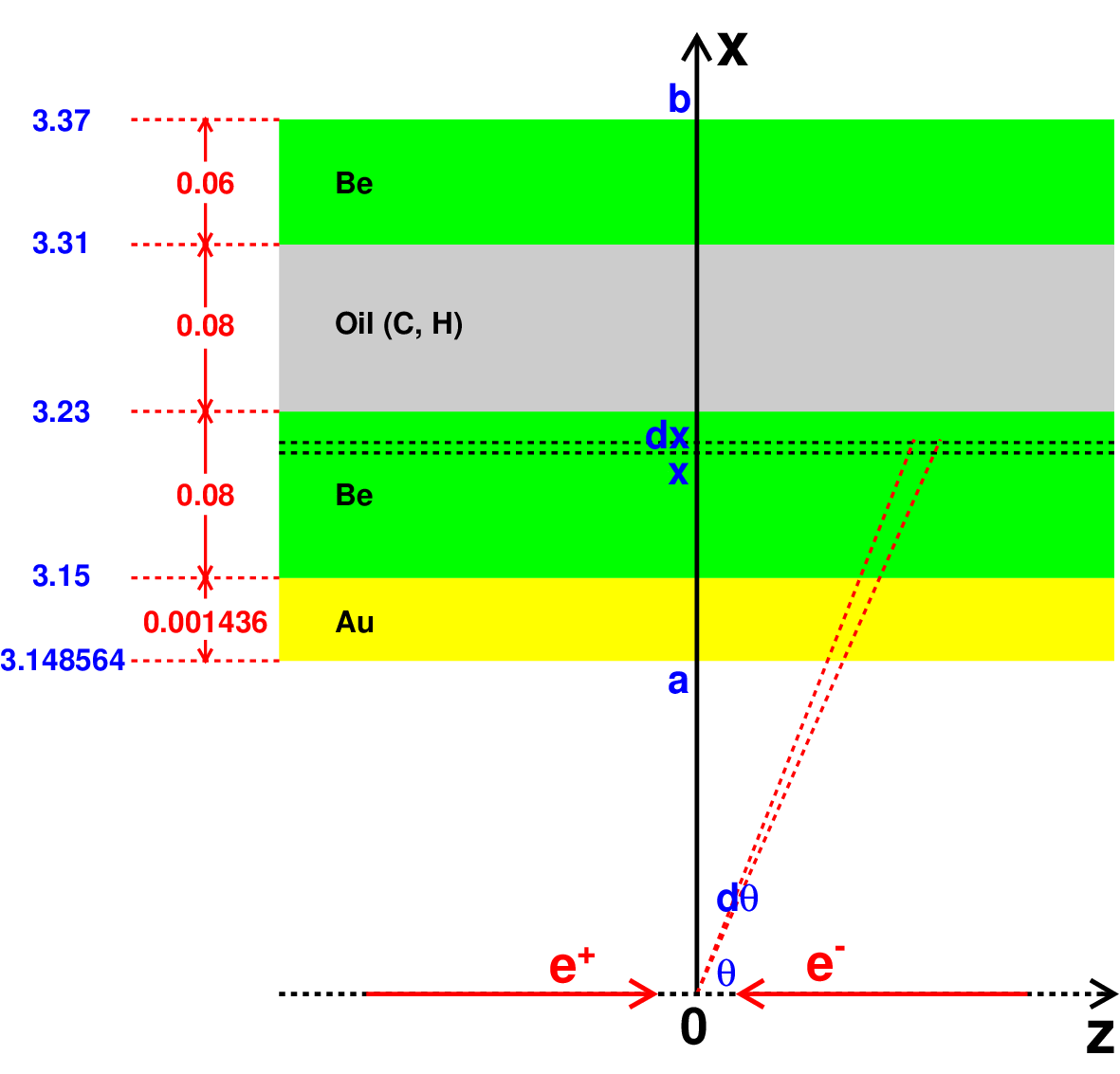}
\end{overpic}
\caption{The schematic diagram of the beam pipe, the length units are centimeter (cm). The $z$-axis is along the $e^+e^-$ beam direction, and the $x$-axis is perpendicular to the $e^+e^-$ beam direction. ``$a$" and ``$b$" are the distances from the inner surface and the outer surface of the beam pipe to the $z$-axis, which are $a=3.148564~\rm{cm}$ and $b=3.37~\rm{cm}$. }
\label{fig:beampipe2}
\end{center}
\end{figure}

The beam pipe is composed of layers of composite material, which is composed of gold ($^{197}\rm{Au}$), beryllium ($^{9}\rm{Be}$) and oil $(^{12}\rm{C}:$$^{1}\rm{H}$$=1:2.13)$, as shown in Fig.~\ref{fig:beampipe2}, and more details can be found in Ref.~\cite{besiii-supp}. The distance from a position to the $z$-axis is defined as $x$, so the number of nuclei ($^{197}\rm{Au}$, $^{9}\rm{Be}$, $^{12}\rm{C}$) per unit volume $N(x)$ is:
\begin{equation}
N(x)=
\begin{cases}
    \frac{\rho_{\rm{Au}}}{A_{\rm{Au}}\cdot 1\rm{u}}=\frac{19.32\rm{g/cm^{3}}}{197\cdot 1.6605\times10^{-27}\rm{kg}}=5.91\times10^{22}~\rm{cm^{-3}}, \ \ \ \ \ \ \ \ \ \ 3.148564\leq \it{x}\leq \rm{3.15}~\rm{cm} \\
    \\
    \frac{\rho_{\rm{Be}}}{A_{\rm{Be}}\cdot 1\rm{u}}=\frac{1.85\rm{g/cm^{3}}}{9\cdot 1.6605\times10^{-27}\rm{kg}}=1.24\times10^{23}~\rm{cm^{-3}}, \ \ \ \ \ \ \ \ \ \ \ \ \ 3.15< \it{x}\leq \rm{3.23}~\rm{cm} \\
    \\
    \frac{\rho_{\rm{Oil}}}{(A_{\rm{C}}+A_{\rm{H}}\cdot 2.13)\cdot 1\rm{u}}=\frac{0.81\rm{g/cm^{3}}}{(12+1\cdot 2.13)\cdot 1\rm{u}}=3.45\times10^{22}~\rm{cm^{-3}}, \ \ \ \ \ 3.23< \it{x}\leq \rm{3.31}~\rm{cm} \\
    \\
    \frac{\rho_{\rm{Be}}}{A_{\rm{Be}}\cdot 1\rm{u}}=\frac{1.85\rm{g/cm^{3}}}{9\cdot 1.6605\times10^{-27}\rm{kg}}=1.24\times10^{23}~\rm{cm^{-3}}, \ \ \ \ \ \ \ \ \ \ \ \ \ 3.31< \it{x}\leq \rm{3.37}~\rm{cm} \\
\end{cases}
\end{equation}
where $\rho$ is the volume density, $A$ is the number of nucleons in the nucleus and u is the atomic mass unit.

There is no definite conclusion about the cross section ratios between the reaction processes $\Xi^0+{^A\rm{X}}\too\Xi^-+p+{^{A-1}\rm{X}}$. Generally, from the measurements of other particle interactions with nuclei, the cross section is proportional to $A^{\alpha'}$, where $A$ is the number of nucleons in the nucleus and $\alpha'$ is an exponential coefficient in the range $\frac{2}{3}$ to $1$~\cite{ratio1-supp, ratio2-supp, ratio3-supp, ratio4-supp, ratio5-supp}. $\alpha'=\frac{2}{3}$ is the most common situation, which corresponds to a pure surface process and the reaction is due to the interaction with single nucleons on the nucleus surface. For the reaction process $\Xi^0+{^A\rm{X}}\too\Xi^-+p+{^{A-1}\rm{X}}$, we assume $\alpha'=\frac{2}{3}$ to get the nominal result of the ratio of cross sections for $^{9}\rm{Be}$, $^{12}\rm{C}$ and $^{197}\rm{Au}$ nuclei, and take $\alpha'=1$ to get the systematic uncertainty. Hence, the cross section with neutron is proportional to $A^{\frac{2}{3}}\times\frac{N}{A}=\frac{N}{A^{\frac{1}{3}}}$, where $A$ and $N$ are the numbers of nucleons and neutrons in the nucleus. Then we can get the cross section ratios as $\sigma^{^{9}\rm{Be}}:\sigma^{^{12}\rm{C}}:\sigma^{^{197}\rm{Au}}=\frac{5}{9^{\frac{1}{3}}}:\frac{6}{12^{\frac{1}{3}}}:\frac{118}{197^{\frac{1}{3}}}
=2.4037:2.6207:20.2796=1.000:1.090:8.437$. We define $\sigma(x)=C(x)\sigma^{^{9}\rm{Be}}=C(x)\sigma(\Xi^0+{^9\rm{Be}}\too\Xi^-+p+{^8\rm{Be}})$, where $C(x)$ is:
\begin{equation}
C(x)=
\begin{cases}
    8.437, \ \ \ \ \ \ \ \ \ \ \ \ \ \ 3.148564~\rm{cm}\leq \it{x}\leq \rm{3.15}~\rm{cm} \\
    \\
    1.000, \ \ \ \ \ \ \ \ \ \ \ \ \ \ 3.15~\rm{cm}< \it{x}\leq \rm{3.23}~\rm{cm} \\
    \\
    1.090, \ \ \ \ \ \ \ \ \ \ \ \ \ \ 3.23~\rm{cm}< \it{x}\leq \rm{3.31}~\rm{cm} \\
    \\
    1.000, \ \ \ \ \ \ \ \ \ \ \ \ \ \ 3.31~\rm{cm}< \it{x}\leq \rm{3.37}~\rm{cm} \\
\end{cases}
\end{equation}

As shown in Fig.~\ref{fig:beampipe2}, at the position of $\theta$ and within the range of ${\rm{d}}\theta$, the number of $\Xi^0$ that can reach the position of $x$ is:
\begin{equation}
    \frac{{\rm{d}}\it{N(\theta)}}{{\rm{d}}\theta}{\rm{d}}\theta e^{-\frac{t}{\tau}\sqrt{1-\frac{v^2}{c^2}}}=\frac{{\rm{d}}\it{N(\theta)}}{{\rm{d}}\theta}{\rm{d}}\theta e^{-\frac{x}{{\rm{sin}}\theta v\tau}\sqrt{1-\frac{v^2}{c^2}}} = \frac{{\rm{d}}N(\theta)}{{\rm{d}}\theta}{\rm{d}}\theta e^{-\frac{x}{\rm{sin}\theta \it{\frac{P_{\rm{\Xi^0}}}{m_{\rm{\Xi^0}}}\tau}}} = \frac{{\rm{d}}N(\theta)}{{\rm{d}}\theta}{\rm{d}}\theta e^{-\frac{x}{\rm{sin}\theta \it{\beta\gamma L}}},
\end{equation}
where $v$ is the speed of $\Xi^0$, $E_{\text{beam}}$ is the $e^+$ or $e^-$ beam energy for data, $\tau$ is mean lifetime of $\Xi^0$, $c$ is speed of light in vacuum, $\beta\gamma\equiv \frac{P_{\rm{\Xi^0}}}{m_{\rm{\Xi^0}}c} = \frac{\sqrt{E_{\rm{beam}}^{\rm{2}}-m_{\rm{\Xi^0}}^{\rm{2}}c^{\rm{4}}}}{m_{\rm{\Xi^0}}c^{\rm{2}}}$, and $L\equiv c\tau$~\cite{pdg-supp}.

Then these $\Xi^0$ particles can interact with neutrons in the material of the beam pipe in the range ${\rm{d}}x$ at position $x$ to produce the reaction process $\Xi^{0}n\too\Xi^{-}p$. So according to the definition of cross section, we get the number of surviving signal events for the reaction process as:
\begin{equation}
    \frac{{\rm{d}}N(\theta)}{{\rm{d}}\theta}{\rm{d}}\theta e^{-\frac{x}{\rm{sin}\theta \it{\beta\gamma L}}} \sigma(x)N(x)\frac{{\rm{d}}x}{\rm{sin}\theta}\epsilon\mathcal{B}=\frac{\it{N}_{\it{J}/\psi}\mathcal{B}_{\it{J}/\psi}}{2+\frac{2}{3}\alpha}(1+\alpha \rm{cos}^2\theta)\it{e}^{-\frac{x}{\rm{sin}\theta \it{\beta\gamma L}}}\sigma(x)N(x)\epsilon\mathcal{B}{\rm{d}}\theta {\rm{d}}x,
\end{equation}
where $\epsilon$ is the selection efficiency, and $\mathcal{B}=\mathcal{B}(\bar{\Xi}^0\too\bar{\Lambda}\pi^0)\mathcal{B}(\bar{\Lambda}\too\bar{p}\pi^+)\mathcal{B}(\pi^0\too\gamma\gamma)
\mathcal{B}(\Xi^-\too\Lambda\pi^-)\mathcal{B}(\Lambda\too p\pi^-)$.

After integrating the above formula in the whole beam pipe region, the total number of surviving signal events $N^{\rm{sig}}$ for the reaction process is:
\begin{equation}
    N^{\rm{sig}}=\int_{a}^{b}\int_{0}^{\pi}\frac{\it{N}_{\it{J}/\psi}\mathcal{B}_{\it{J}/\psi}}{2+\frac{2}{3}\alpha}(1+\alpha \rm{cos}^2\theta) \it{e}^{-\frac{x}{\rm{sin}\theta \it{\beta\gamma L}}}\sigma(x)N(x)\epsilon\mathcal{B}{\rm{d}}\theta {\rm{d}}x.
\end{equation}
Here, the beam pipe can be regarded as infinitely long with respect to the product $\beta\gamma L$ of $\Xi^0$.

Therefore, the cross section formula of $\Xi^0+{^9\rm{Be}}\too\Xi^-+p+{^8\rm{Be}}$ is:
\begin{equation}
    \sigma(\Xi^0+{^9\rm{Be}}\too\Xi^-+p+{^8\rm{Be}}) = \frac{N^{\rm{sig}}} {\epsilon \mathcal{B} \frac{\it{N}_{\it{J}/\psi}\mathcal{B}_{\it{J}/\psi}}{\rm{2}+\frac{\rm{2}}{\rm{3}}\alpha} \int_{a}^{b}\int_{\rm{0}}^{\pi} (\rm{1}+\alpha \rm{cos}^2\theta) \it{e}^{-\frac{x}{\rm{sin}\theta \it{\beta\gamma L}}}N(x)C(x) \rm{d}\theta \rm{d}\it{x}},
\end{equation}

\section{II. systematic uncertainties}
The sources of systematic uncertainties related to the measured cross sections $\sigma(\Xi^0+{^9\rm{Be}}\too\Xi^-+p+{^8\rm{Be}})$ and $\sigma(\Xi^0+{^{12}\rm{C}}\too\Xi^-+p+{^{11}\rm{C}})$ are discussed below. The uncertainty in the tracking efficiency, photon efficiency, and PID efficiency is $1\%$ per track or per photon~\cite{track-supp}. The uncertainty from the track number requirement is studied by the control sample $J/\psi\too\Xi^-\bar{\Xi}^+\too\Lambda\pi^-\bar{\Lambda}\pi^+\too p\pi^-\pi^-\bar{p}\pi^+\pi^+$. We enlarge the nominal mass windows and $R_{xy}$ requirement by $20\%$ to compare the difference of the result to the nominal one. The requirement on the $M_{\rm{recoil}}(\bar{\Xi}^0\Lambda)$ is changed to less than $0.1$~GeV/$c^2$ to estimate the uncertainty.

In the MC simulation, we take the momentum of the neutron in the nucleus as zero, but due to the existence of the Fermi-momentum, there is a difference in the distribution of $(\Xi^-+p)$ momentum $P(\Xi^-+p)$ for data and MC. The monoenergetic momentum of the incident $\Xi^0$ is very high compared with the Fermi-momentum, according to the rule of momentum synthesis, the change of the $P(\Xi^-+p)$ for most events is within $\pm0.1$~GeV/$c$. So to estimate the uncertainty from $P(\Xi^-+p)$, we vary the momentum of the free neutron by $\pm0.1$~GeV/$c$ along the direction of the incident $\Xi^0$ in the generated signal MC, and take the larger difference as the uncertainty. We assume the distribution of $M(\Xi^-p)$ is flat in the nominal signal MC. To get the uncertainty from the $M(\Xi^-p)$ distribution, the difference in the efficiency between the nominal signal MC and the weighted MC according to the distribution of signal events in data is taken as systematic uncertainty. The reaction $\Xi^{0}n\too\Xi^{-}p$ is simulated with a uniform angular distribution over the phase-space to estimate the nominal efficiency. The weighted efficiency of signal events is calculated based on real data, as shown in Fig.~\ref{fig:fig_angle}. The difference between the nominal efficiency and weighted efficiency is taken as the uncertainty from the angular distribution.
\begin{figure}[htbp]
\begin{center}
\begin{overpic}[width=0.34\textwidth]{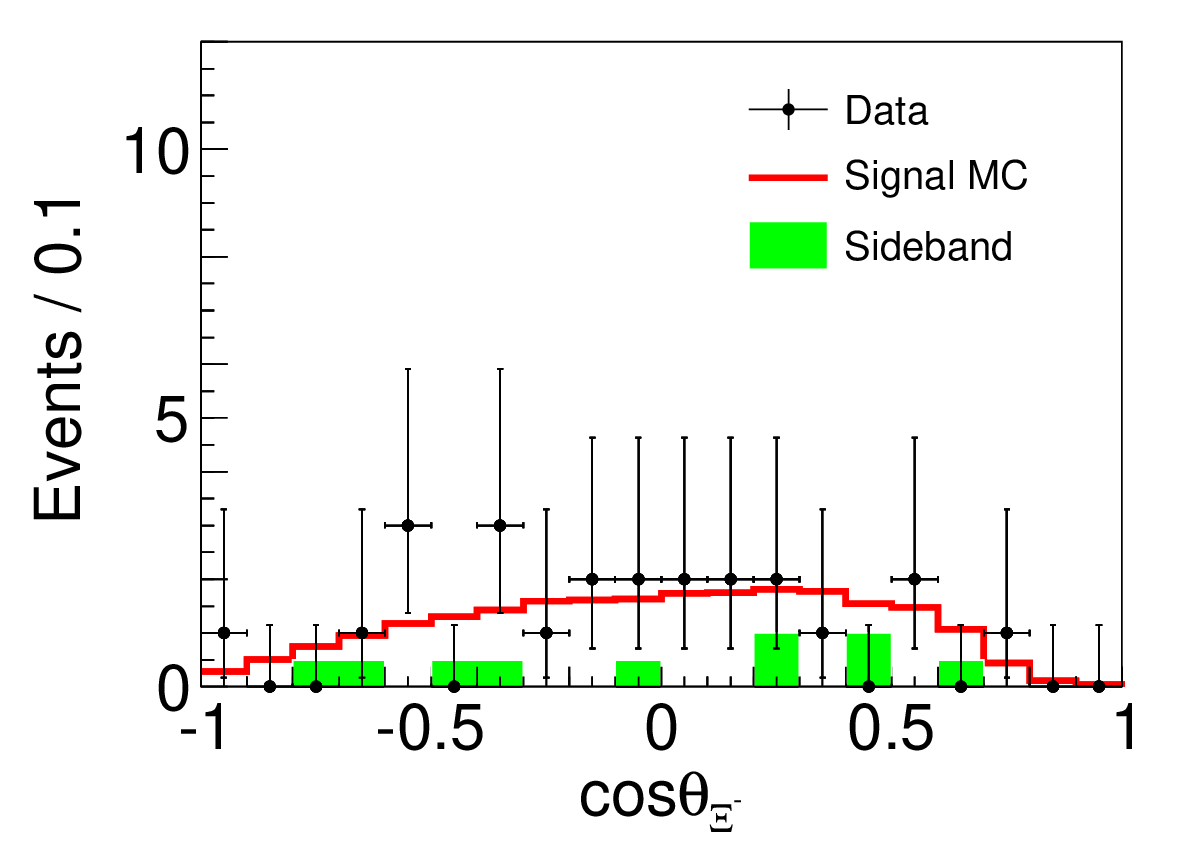}
\end{overpic}
\caption{Distribution of cos$\theta_{\Xi^-}$ for data, where $\theta_{\Xi^-}$ is the scattering angle of the $\Xi^-$ in the $\Xi^-p$ rest frame. The green-shaded histogram corresponds to the normalized events from the $\Xi^-$ sideband region, and the signal MC distribution is normalized by the total number of events for data.}
\label{fig:fig_angle}
\end{center}
\end{figure}

The uncertainty from the MC statistics is estimated according to the number of generated signal MC events. To estimate the uncertainty from the efficiency curve parameterization, we replace the constant with a first-order polynomial function to parameterize the efficiency curve in the beam pipe region, and the change in the results is taken as systematic uncertainty. The uncertainty from the fit procedure includes the signal shape, the fit range and the background shape. The uncertainty from the signal shape is estimated by using the MC-determined signal shape convolved with a free Gaussian function to instead and compare the difference, the uncertainty from the fit range is obtained by varying the limit of the fit range by $\pm10~\text{MeV}/c^{2}$, and the uncertainty associated with the background shape is estimated by changing a first-order polynomial function to a second-order one or a constant.

The uncertainty from the number of $J/\psi$ events is estimated in Ref.~\cite{totalnumber-supp}, and the uncertainty of the branching fractions is taken from the PDG~\cite{pdg-supp}. To estimate the uncertainties from the angular distribution of $J/\psi\too\Xi^0\bar{\Xi}^0$ and the $\Xi^0$ mean lifetime, we vary the angular distribution parameter $\alpha$ and the mean lifetime by $\pm1\sigma$. The uncertainty from the position of the $\EE$ interaction point is obtained by changing the integral range by $\pm0.1$~cm, which is from $(a, b)$ to $(a+0.1, b+0.1)$ or $(a-0.1, b-0.1)$, and the larger difference in the result is taken as the uncertainty. Because the beam pipe is made up of composite material, to extract the cross section $\sigma(\Xi^0+{^9\rm{Be}}\too\Xi^-+p+{^8\rm{Be}})$ or $\sigma(\Xi^0+{^{12}\rm{C}}\too\Xi^-+p+{^{11}\rm{C}})$, we assume the reaction is due to the interaction with single neutrons on the nucleus surface. To estimate the uncertainty from the assumption of the cross section ratio, we choose an extreme assumption that the cross section is proportional to the number of neutrons in the nucleus ($\alpha'=1$), and the difference in the result for the two different extreme assumptions is taken as the uncertainty.

A summary of the systematic uncertainties is presented in Table~\ref{tab:sumerror}, and the total systematic uncertainty is obtained by adding all the individual components in quadrature.
\begin{table}[htbp]
\begin{center}
\caption{Summary of systematic uncertainties (in $\%$).}
\label{tab:sumerror}
\begin{tabular}{ ccc}
  \hline
  \hline
  \ \ Source \ \ & \ \ $\sigma(\Xi^0+{^9\rm{Be}}\too\Xi^-+p+{^8\rm{Be}})$ \ \ & \ \ $\sigma(\Xi^0+{^{12}\rm{C}}\too\Xi^-+p+{^{11}\rm{C}})$ \ \  \\
  \hline
  Tracking efficiency                                       & 6.0 & 6.0 \\
  Photon efficiency                                      & 2.0 & 2.0 \\
  PID efficiency                                         & 6.0 & 6.0 \\
  Track number                                           & 3.0 & 3.0 \\
  Mass windows                                           & 7.8 & 7.8 \\
  $R_{xy}$ requirement                                   & 6.6 & 6.6 \\
  $M_{\text{recoil}}(\bar{\Xi}^0\Lambda)$ requirement    & 4.3 & 4.3 \\
  $(\Xi^-+p)$ momentum                                   & 10.0 & 10.0 \\
  $M(\Xi^-p)$ distribution                               & 0.6 & 0.6 \\
  Angular distribution of $\Xi^{0}n\too\Xi^{-}p$         & 1.0 & 1.0 \\
  MC statistics                                          & 0.7 & 0.7 \\
  Efficiency curve parameterization                      & 0.5 & 0.5 \\
  Fit procedure                                          & 5.0 & 5.0 \\
  Number of $J/\psi$                                     & 0.4 & 0.4 \\
  Branching fractions                                    & 3.6 & 3.6 \\
  Angular distribution of $J/\psi\too\Xi^0\bar{\Xi}^0$   & 0.1 & 0.1 \\
  $\Xi^0$ mean lifetime                                  & 2.7 & 2.7 \\
  $e^+e^-$ interaction point                             & 2.7 & 2.7 \\
  Cross section ratios                                   & 7.2 & 2.1 \\
  \hline
  sum                                                    & 20.4 & 19.2 \\
  \hline
  \hline
\end{tabular}
\end{center}
\end{table}

\end{widetext}

\end{document}